# Efficacy Analysis of Power Swing Blocking and Out-of-Step Tripping Functions in Grid-Following VSC Systems

Yongxin Xiong, Heng Wu, *Member, IEEE*, Xiongfei Wang, *Fellow, IEEE*, and Yifei Li, *Student Member, IEEE*

*Abstract*—Power flow oscillations can occur in power systems after major disturbances such as system faults, which may result in significant power swings and even lead to system collapse. This paper provides a detailed analysis of the efficacy of the legacy power swing blocking and out-of-step tripping functions in grid-following voltage-source converter (GFL-VSC) systems. First, the power swing dynamics of GFL-VSC are characterized, considering the outer power control loops and phase-locked loop (PLL). It is found that the loss of synchronism (LOS) of the GFL-VSC system is caused by the divergence of the output angle of the PLL, which is fundamentally different from that of synchronous generator (SG)-based systems, whose LOS originates from the divergence of power angle differences between power sources. Therefore, the legacy setting principles for power swing blocking and out-of-step tripping function designed for SG-based systems can not be directly applied to GFL-VSC systems, and may even malfunction. Time-domain simulations are performed in the PSCAD/EMTDC platform to validate the correctness of the theoretical analysis.

*Index Terms*—VSC-connected system, grid-following control, PLL-synchronization, power swing blocking, out-of-step tripping

## I. INTRODUCTION

The integration of renewable energy sources into power grids keeps increasing in recent years. Voltage-source converters (VSCs) are commonly found in renewable energy sources, where the grid-following (GFL) control is still the prevalent control scheme used with VSCs. GFL-VSCs use the phase-locked loop (PLL) to track the phase and frequency at the point of common coupling (PCC) of the ac grid [1], [2], and they may oscillate with other GFL-VSCs/synchronous generators (SGs) under large disturbances, manifesting stable or unstable power swings [3], [4]. With stable power swings, the oscillation is finally damped out and the system returns to an equilibrium state after disturbances. In contrast, unstable power swings may lead to large power fluctuations and trigger unnecessary relay operations, causing the loss of synchronism (LOS) and even power outages [4]-[6].

To minimize the impact of power swings on the reliability of the power system, the power swing blocking (PSB) and out-of-step tripping (OST) functions are introduced in the legacy protection relays [7]. The PSB function is used to discriminate between system faults and power swings and block the related protection relays, especially the distance protection relays, to prevent unnecessary disconnections during the stable power swing [7]. The OST function is used to discriminate between stable and unstable power swings. If unstable power swings are detected, the OST function will take effect and trigger the protection relays [8]-[11].

The implementations of PSB and OST require the accurate detection of stable/unstable power swings. In SG-based power systems, the stable/unstable power swings are dictated by the convergence/divergence of power angle differences between different power sources. Further, the dynamics of this power angle difference can be reflected in the impedance trajectory of SG. Hence, checking the impedance trajectory of SG under the dual-blinder scheme is a common practice for power swing detection [12], [13]. With the proper configurations of outer, middle, and inner blinders, the time intervals between the impedance trajectory of SGs that cross different blinders are measured and compared with different thresholds to detect stable/unstable power swings [2], [13]. It is worth mentioning that the settings of these blinders and time thresholds are based on the power swing dynamics of SGs, which, however, cannot be directly extended to the GFL-VSC system, since the power swing dynamics of GFL-VSCs are governed by the used power and synchronization control algorithms.

Many alternative power-swing detection methods, such as the Taylor Series expansion-based method [14], the Lissajous Figure-based method [15], the rate of impedance angle change-based method [16], the moving window averaging method [17], and the wavelet transform-based scheme [18], are also reported in the literature. However, those methods are still based on the dynamics of SGs.

In recent years, some studies have reported on the efficacy of legacy power swing protection schemes in VSC systems. It is pointed out in [2] and [3] that the large-scale integration of wind power generation can change power swing dynamics and may cause the maloperation of legacy power swing protection schemes. While possible solutions are reported in [2] and [3] to ensure effective power swing detections, the studies are all case-specific and lack analytical insight, and the wind power generators are merely represented as PQ sources without considering control impacts of VSCs. The impacts of different short-circuit characteristics of VSCs on the maloperations of legacy PSB and OST functions are discussed in [19], yet the conclusion is also drawn case by case, and analytical insights

Yongxin Xiong, Heng Wu, and Yifei Li are with the AAU Energy, Aalborg University, 9220 Aalborg, Denmark (e-mail: yxio@energy.aau.dk; hew@energy.aau.dk; xwa@energy.aau.dk; yili@energy.aau.dk).

Xiongfei Wang is with KTH Royal Institute of Technology, Stockholm, Sweden. (email: xiongfei@kth.se).

Color versions of one or more of the figures in this article are available online at http://ieeexplore.ieee.org

into the control impacts of VSCs are still absent.

To fill the void, this paper provides an in-depth theoretical analysis of the impact of GFL control on the efficacy of the legacy PSB and OST functions in GFL-VSC systems. Without loss of generality, the widely used dual-blinder scheme is adopted as a benchmark for power swing detection. The main contributions of this paper are as follows:

- The relationship among the impedance trajectory of GFL-VSC, the output phase angle of the PLL in reference to the grid phase (i.e. $\delta_{PLL}$), and the phase angle of the current vector that is dictated by the outer active and reactive power control loop (i.e. $\varphi$) are established in this work. It is found that the power swing dynamics of GFL-VSC are highly control dependent and characterized by both $\delta_{PLL}$ and $\varphi$, which are fundamentally different from the power swing dynamics of SGs that are determined by the physical angle difference between power sources.
- It is also revealed that the phase angle $\varphi$ essentially offsets the influence of $\delta_{PLL}$ during the dynamic process. This offset could suppress the power swings and limit the impedance trajectory to a small range, making it difficult to detect both stable and unstable power swings.
- As the impedance trajectory can be affected by control-dependent phase angles, i.e., $\delta_{PLL}$ and $\varphi$, in GFL-VSCs, the dual-blinder scheme that uses impedance trajectory to predict physical power angle dynamics in SG-based systems cannot be directly used with GFL-VSC systems.

The rest of the paper is organized as follows. First, the widely used dual-blinder schemes for the PSB and OST functions are introduced in Section II. Then, in Section III, the efficacy of PSB and OST functions in VSC-connected systems with GFL control is analyzed theoretically. The validity of the theoretical analysis is confirmed through simulations in the PSCAD/EMTDC platform in Section IV. Finally, conclusions are drawn in Section V.

## II. POWER SWING DETECTION METHOD

This section provides a brief review of the power swing phenomenon and the principle of the dual-blinder scheme used in SG-based systems [2], [4]. Among different power swing detection methods, the dual-blinder scheme based on the rate of change of impedance is widely used in practice. In this section, it is employed as a benchmark to analyze the performance of PSB and OST functions in GFL-VSC systems.

### A. Impedance Trajectory during Power Swing

The two-source system shown in Fig. 1 (a) is introduced to elaborate on the power swing phenomenon in SG-based systems, whose phasor diagram is shown in Fig. 1 (b). In Fig. 1 (a), the left source has a phase angle leading to the right source (i.e. $\delta_S$). During power swings, $\delta_S$ varies with time. $Z_S\angle\theta_S$ and $Z_R\angle\theta_R$ denote the equivalent output impedance of two sources, respectively. $Z_L\angle\theta_L$ denotes the line impedance. The total equivalent impedance $Z_T\angle\theta_T$ can be derived as $Z_T\angle\theta_T = Z_S\angle\theta_S + Z_L\angle\theta_L + Z_R\angle\theta_R$.

The voltage and current at the Point of Common Coupling (PCC) as measured by protection relay $RL_1$, where $\varphi$ denotes the angle difference between the PCC voltage and the system current in SG-based systems. The impedance measured at the PCC by the relay $RL_1$, i.e., $Z_{PCC}\angle\theta_{ZP}$ can be expressed as [4]:

$$Z_{PCC}\angle\theta_{ZP} = \frac{V_{PCC}\angle\delta_{PCC}}{I_g\angle(\delta_{PCC}-\varphi)}$$
$$= Z_T\angle\theta_T \cdot \frac{E_S\angle\delta_S}{E_S\angle\delta_S - E_R} - Z_S\angle\theta_S \quad (1)$$

**Fig. 1.** (a) A typical two-source power system. (b) Phasor diagram of the two-source system.

**Fig. 2.** Impedance trajectory of $Z_{PCC}$ measured by $RL_1$ during power swings with different source voltage magnitude ratios.

Fig. 2 illustrates the impedance trajectory of the two-source system during power swings. The vector **OP** denotes $Z_{PCC}\angle\theta_{ZP}$, while the vectors **AP** and **BP** represent the $E_R\angle\delta_S/I_g\angle(\delta_{PCC}-\varphi)$ and $E_R\angle0/I_g\angle(\delta_{PCC}-\varphi)$, respectively. It can be found that the angle difference between the **AP** and **BP** is equal to the angle difference between $E_S$ and $E_R$, i.e., $\delta_S$. Thus, with $\delta_S$ varying during power swings, the impedance trajectory of $Z_{PCC}\angle\theta_{ZP}$ can be employed to predict the variations of $\delta_S$, and then characterize the power swing [4]. With different operating conditions, the ratio of $E_S$ over $E_R$ may appear different, influencing the impedance trajectories. Defining $n$ as the ratio of $E_S$ over $E_R$, $n$ can be expressed as follows:

$$n = |\frac{E_S}{E_R}| = \frac{PA}{PB} \quad (2)$$

*Condition 1: n=1.* The magnitude of $Z_{PCC}$ can be calculated as [4]:

$$Z_{PCC} = Z_T \cdot \frac{\angle\delta_S}{\angle\delta_S - 1} - Z_S = \frac{Z_T}{2} \cdot (1 - j\cot\frac{\delta_S}{2}) - Z_S \quad (3)$$

It can be found from (3) that with the variations of the $\delta_S$,

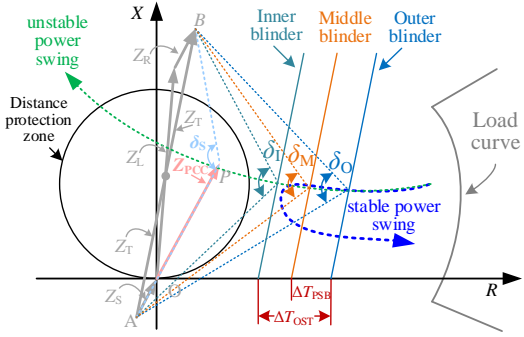

**Fig. 3.** Dual-blinder schemes for PSB and OST.

the trajectory of $Z_{PCC}$ can be represented as a straight line intersecting the segment $Z_T$ at its midpoint, as shown in Fig. 2.

*Condition 2: n≠1.* If $n>1$, the $Z_{PCC}$ trajectory is above the straight trajectory of $n=1$, while if $n<1$, the $Z_{PCC}$ trajectory is under the straight line as shown in Fig. 2 [4]. In both cases, when the phase angle $\delta_S$ increases, the impedance trajectory (red dash line in Fig. 2) shifts towards the left plane, increasing the risk of LOS.

*B. Dual-blinder Scheme for Power Swing Detection*

As shown in Fig. 2, the power swings can be reflected by the impedance trajectory. For stable power swings, the used protection elements, e.g. the distance protection relay, should be blocked to avoid maloperation, which is known as the PSB function. For unstable power swings, it is used to discriminate between stable and unstable power swings. If unstable power swings are detected, the OST function will take effect and trigger the protection relays.

The dual-blinder scheme is a commonly used method for power swing detection, which is based on the rate of change of impedance trajectory [2], [4]. With the proper configurations of outer, middle, and inner blinders, the time intervals between the blinders that impedance trajectory crosses are measured and compared with the thresholds to detect stable/unstable power swings in Fig. 3 [2], [4]. The details of PSB and OST functions are introduced as follows.

*1) PSB function:* The time interval $\Delta t_1$ for the impedance trajectory crossing the outer and middle blinders is measured. If $\Delta t_1$ is larger than a setting value $\Delta T_{PSB}$, a power swing is detected. Conversely, a system fault is declared [2], [4].

*2) OST function:* The time interval $\Delta t_2$ for the impedance trajectory crossing the outer and inner blinders is measured. If $\Delta t_2$ is larger than a setting value $\Delta T_{OST}$, an unstable power swing will be declared. Otherwise, a stable power swing will be declared [2], [4].

The setting principles of different blinders are elaborated in detail in [2] and [4], which are introduced as follows.

*1) Outer Blinder: setting away from the maximum expected line loading.* When $\delta_S$ is equal to 90º in Fig. 1, the maximum load transfer between two sources is achieved. To make sure the power swing detection does not be triggered during heavy load situations, the angle of the Outer Blinder (i.e., $\delta_O$ in Fig. 3) is recommended to be set to 90° according to [4].

*2) Middle Blinder: setting outside the most overreaching protection zone that is to be blocked when a power swing*

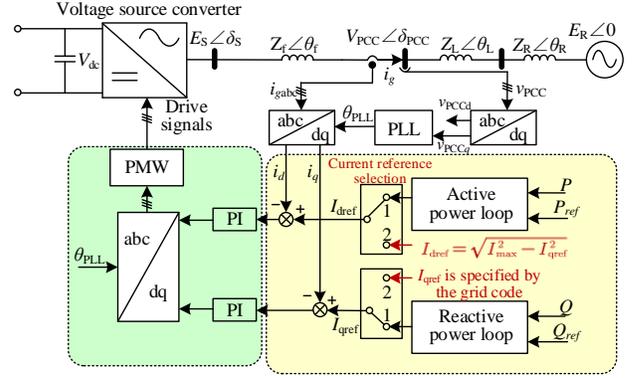

**Fig. 4.** Single-line diagram of the GFL-VSC system with the fault ride-through control.

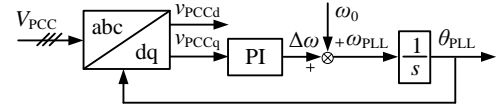

**Fig. 5.** Block diagram of SRF-PLL.

*condition occurs (e.g., Zone 1 in distance protection).* The angle of the Middle Blinder (i.e. $\delta_M$ in Fig. 3), should be configured based on the size of the protection zone [4].

*3) Inner Blinder: ensuring that stable power swings will never cross it.* The Inner Blinder angle (i.e. $\delta_I$ in Fig. 3) should be set larger than the critical angle of the power system, and $\delta_I$ is recommended to be set to 120° according to [4].

*4) Time interval settings:* $\Delta T_{PSB}$ and $\Delta T_{OST}$ are calculated as [2], [4]:

$$\Delta T_{PSB} = \frac{(\delta_M - \delta_O) \cdot f_0}{360° \cdot f_{swing}} \quad \text{(cycles)} \quad (4)$$

$$\Delta T_{OST} = \frac{(\delta_I - \delta_O) \cdot f_0}{360° \cdot f_{swing}} \quad \text{(cycles)} \quad (5)$$

where $\delta_O$, $\delta_M$, and $\delta_I$ represent the angles when the impedance trajectory crosses the Outer, Middle, and Inner Blinders depicted in Fig. 3, respectively. $f_0$ refers to the normal frequency of the power system. $f_{swing}$ represents the slip frequency of the power swing [4].

## III. EFFICACY ANALYSIS OF PSB AND OST FUNCTIONS IN GFL-VSC SYSTEMS

This section provides an in-depth theoretical analysis of the efficacy of PSB and OST functions in GFL-VSC systems.

*A. GFL-VSC System under Study*

Fig. 4 illustrates the single-line diagram of the GFL-VSC system. $Z_f \angle \theta_f$ and $Z_L \angle \theta_L$ represent the impedance of the converter output filter and the transmission line, respectively. The phase angle between the voltage of the VSC and the equivalent source (i.e., $E_S$ and $E_R$) is denoted as $\delta_S$. $V_{PCC}$ denotes the amplitude of the PCC voltage, while $\delta_{PCC}$ is the phase angle difference between the PCC voltage and grid voltage. The three-phase PCC voltage can be expressed as:

$$\begin{cases} V_{PCCa} = V_{PCC}\cos(\omega t + \delta_{PCC}) \\ V_{PCCb} = V_{PCC}\cos(\omega t + \delta_{PCC} - 120°) \\ V_{PCCc} = V_{PCC}\cos(\omega t + \delta_{PCC} + 120°) \end{cases} \quad (6)$$

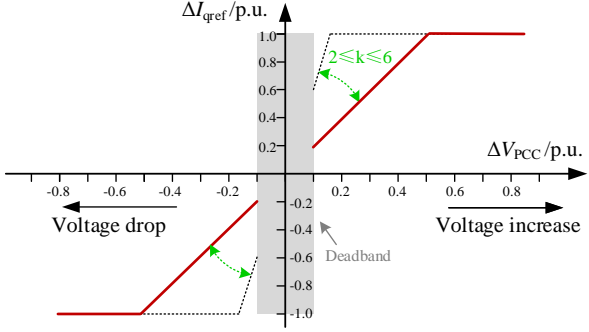

**Fig. 6.** German grid code on the reactive current injection during fault ride-through operations.

where $V_{PCCa}$, $V_{PCCb}$, and $V_{PCCc}$ denote the abc-frame components of $V_{PCC}$, respectively. In Fig. 4, $I_{dref}$ and $I_{qref}$ are references for the active and the reactive currents, respectively. The synchronous reference frame (SRF)-PLL is employed for synchronizing GFL-VSC with the power grid, whose control diagram is depicted in Fig. 5. $\omega_0$ represents the nominal grid frequency, $\Delta\omega$ and $\omega_{PLL}$ represent the frequency deviation and generated frequency by PLL, respectively. $K_{pP}$ and $K_{iP}$ represent the proportional and integral gain of the PI controller of the PLL. $\theta_{PLL}$ represents the output phase angle of the PLL [20]. The three-phase voltage at the PCC of VSC, i.e., $V_{PCC}$ is measured by the PLL and then transformed to the $dq$ frame as:

$$\begin{bmatrix} v_{PCCd} \\ v_{PCCq} \end{bmatrix} = \frac{2}{3} \begin{bmatrix} \cos\theta_{PLL} & \cos(\theta_{PLL}-120°) & \cos(\theta_{PLL}+120°) \\ \sin\theta_{PLL} & \sin(\theta_{PLL}-120°) & \sin(\theta_{PLL}+120°) \end{bmatrix} \begin{bmatrix} V_{PCCa} \\ V_{PCCb} \\ V_{PCCc} \end{bmatrix} \quad (7)$$

where $v_{PCCd}$ and $v_{PCCq}$ denote the dq-frame components of $V_{PCC}$, respectively. Based on Fig. 5, the dynamic model of PLL is given by:

$$\theta_{PLL} = \int [\omega_0 + (K_{pP} + \int K_{iP}) \cdot v_{PCCq}] \, dt \quad (8)$$

The q-axis voltage $v_{PCCq}$ is further regulated by using a PI controller. In the steady state, $v_{PCCq}=0$ and $\theta_{PLL}=\omega_0 t+\delta_{PCC}$. Defining the angle difference between $\theta_{PLL}$ and $E_R$ as $\delta_{PLL}$, $\delta_{PLL}$ can be calculated as:

$$\delta_{PLL} = \theta_{PLL} - \theta_{E_R} = \theta_{PLL} - \omega_0 t \quad (9)$$

It should be noted that $\delta_{PLL}=\delta_{PCC}$ at the steady state. Substituting (6) and (9) into (7), $v_{PCCdq}$ can be derived as:

$$\begin{cases} v_{PCCd} = V_{PCC}\cos(\delta_{PCC} - \delta_{PLL}) \\ v_{PCCq} = V_{PCC}\sin(\delta_{PCC} - \delta_{PLL}) \end{cases} \quad (10)$$

*1) Fault ride-through control.* According to grid codes in [21] and [22], the VSC should inject the reactive current based on the variations of $V_{PCC}$ during larger disturbances, as shown in Fig. 6. The $k$-factor is chosen to 2 in this work. According to Figs. 4 and 6, during normal operation, the current reference selection is switched to 1, where $I_{dref}$ and $I_{qref}$ are determined by the active and reactive power control loop, respectively. During grid faults, the current reference selection is switched to 2, where $I_{qref}$ is specified directly based on the grid code requirements, and $I_{dref}$ is adjusted based on $I_{qref}$ to prevent overcurrent during fault ride-through [20].

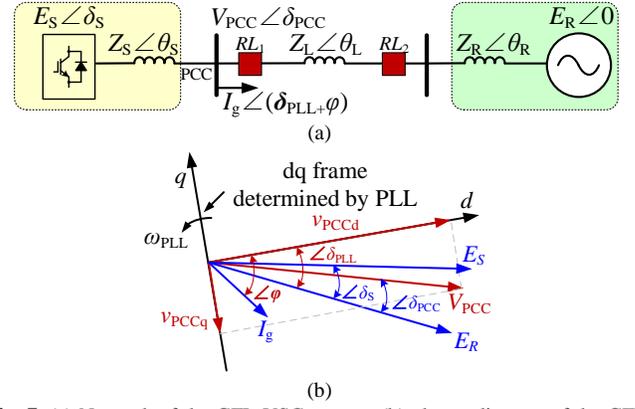

**Fig. 7.** (a) Network of the GFL-VSC system, (b) phasor diagram of the GFL-VSC system.

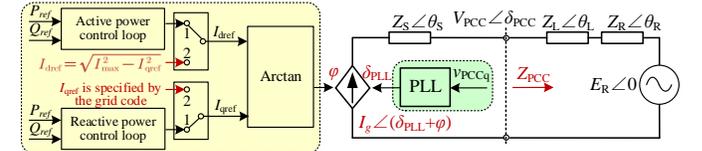

**Fig. 8.** Simplified diagram of the GFL-VSC system.

*2) Inner-loop current control.* The $dq$-frame current control with PI regulators is adopted. Since the bandwidth of the inner loop current control (usually hundreds of Hz) is much higher than that of the outer-loop power control and PLL, the inner current loop can be regarded as a unity gain [20].

*3) Outer-loop power control.* The open-loop active and reactive power control is employed as the outer loop control in this paper, with which, $I_{dref}$ and $I_{qref}$ can be calculated as [23]:

$$\begin{cases} I_{dref} = \dfrac{v_{PCCd}}{\sqrt{v_{PCCd}^2 + v_{PCCq}^2}} P_{ref} - \dfrac{v_{PCCq}}{\sqrt{v_{PCCd}^2 + v_{PCCq}^2}} Q_{ref} \\ I_{qref} = \dfrac{v_{PCCq}}{\sqrt{v_{PCCd}^2 + v_{PCCq}^2}} P_{ref} + \dfrac{v_{PCCd}}{\sqrt{v_{PCCd}^2 + v_{PCCq}^2}} Q_{ref} \end{cases} \quad (11)$$

where $P_{ref}$ and $Q_{ref}$ represent the reference values of the open-loop active and reactive power control.

### B. Impact of GFL Control on PSB and OST

The equivalent circuit of the GFL-VSC system is illustrated in Fig. 7 (a), and its phasor diagram is depicted in Fig. 7 (b). As the inner current control loop is regarded as a unity gain, the equivalent circuit is further simplified as a controlled current source [20], as shown in Fig. 8. The magnitude and phase angle of the current source are denoted by $I_g$ and $\varphi+\delta_{PLL}$, respectively, where $\delta_{PLL}$ is determined by the dynamics of the PLL, and $\varphi$ is determined by the current reference, i.e.,

$$\varphi = \arctan(\frac{i_q}{i_d}) \approx \arctan(\frac{I_{qref}}{I_{dref}}) \quad (12)$$

Based on Fig. 8, the measured voltage at the PCC of VSC can be expressed as the sum of grid voltage and the voltage drop across the line impedance, which is given by

$$V_{PCC}\angle\delta_{PCC} = E_R\angle 0 + I_g\angle(\varphi + \delta_{PLL}) \cdot (Z_L\angle\theta_L + Z_R\angle\theta_R) \quad (13)$$

Based on (13), using $Z_g$ and $\theta_g$ to represent the magnitude and phase of the grid impedance, i.e., $Z_g\angle\theta_g = Z_L\angle\theta_L+Z_R\angle\theta_R$,

$Z_{\text{PCC}}$ can be further expressed by:

$$Z_{\text{PCC}} \angle \theta_{\text{ZP}} = \frac{V_{\text{PCC}} \angle \delta_{\text{PCC}}}{I_{\text{g}} \angle (\varphi + \delta_{\text{PLL}})} \\ = \frac{E_{\text{R}}}{I_{\text{g}}} \angle -(\varphi + \delta_{\text{PLL}}) + Z_{\text{g}} \angle \theta_{\text{g}} \quad (14)$$

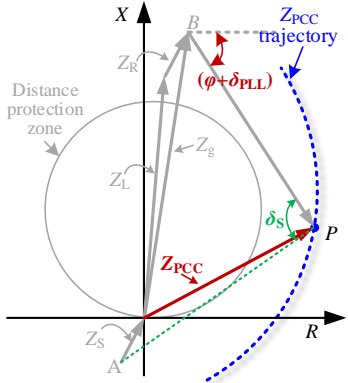

Fig. 9. Impedance trajectory measured at the PCC in GFL-VSC systems.

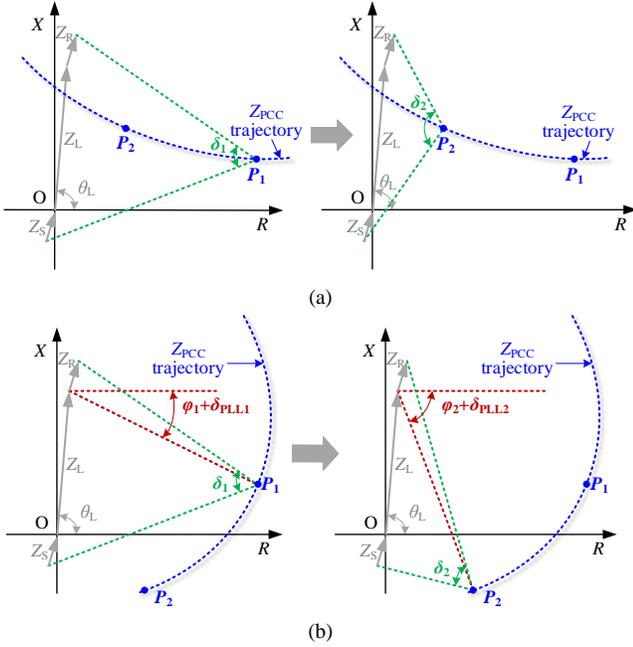

(a)

(b)

Fig. 10. Impedance trajectory predicted by different angles. (a) SG-based systems, (b) GFL-VSC systems.

where $E_{\text{R}}$ is the voltage magnitude of the equivalent power source, which remains constant during power swing situations. $I_{\text{g}}$ is the grid current magnitude, which also remains constant during power swing dynamics, due to the current limitation of GFL-VSC. Therefore, with constant $E_{\text{R}}$, $I_{\text{g}}$, and $Z_{\text{g}} \angle \theta_{\text{g}}$, the impedance trajectory of measured impedance at the PCC, i.e. $Z_{\text{PCC}}$, is dependent on the dynamics of $\delta_{\text{PLL}}+\varphi$, which forms a circle in the complex plane with the radius of $E_{\text{R}}/I_{\text{g}}$ as shown in Fig. 9. With $\varphi+\delta_{\text{PLL}}$ increasing, the impedance trajectory could move clockwise towards the left plane.

*1) Constant φ:* To provide a better analytical insight, a constant power factor control mode is first investigated by assuming a constant $\varphi$ (the impact of active/reactive power control on the dynamics of $\varphi$, and subsequently on the impedance trajectory will be addressed in the next part). In this scenario, it is known from (14) that the dynamics of $Z_{\text{PCC}}$ are solely determined by $\delta_{\text{PLL}}$. A graphical comparison of the impedance trajectories between SG and GFL-VSC during the power swing is given in Fig. 10, considering that the operation points move from $P_1$ to $P_2$, from which, two fundamental differences can be identified:

- In conventional SG-based systems, the internal voltage magnitude of SG, i.e., $E_{\text{S}}$ remains relatively constant during a power swing, causing the impedance trajectory of SG to move leftwards as $\delta_{\text{S}}$ increases. As depicted in Fig. 10 (a), the impedance trajectory moves from operating point $P_1$ to $P_2$ as $\delta_{\text{S}}$ increases from $\delta_1$ to $\delta_2$. In contrast, in the GFL-VSC system, the magnitude of the internal voltage of VSC ($E_{\text{S}}$) may have a significant change in order to maintain the constant magnitude of the output current ($I_{\text{g}}$). Therefore, the leftward movement of the impedance trajectory does not necessarily indicate an increase of $\delta_{\text{S}}$ in such systems. As shown in Fig. 10 (b), the impedance trajectory of the GFL-VSC system also moves leftwards from operating point $P_1$ to $P_2$, yet very limited change can be observed in $\delta_{\text{S}}$.

- Despite the accurate prediction of $\delta_{\text{S}}$ in GFL-VSC systems, it may not reflect the swing dynamics of GFL-VSC. As pointed out in [20], the unstable power swing of GFL-VSC is characterized by the divergence of $\delta_{\text{PLL}}$ that is generated by the PLL. However, the divergence of $\delta_{\text{PLL}}$ does not necessarily correlate to the divergence of the physical power angle $\delta_{\text{S}}$. As shown in Fig. 10 (b), while $\delta_{\text{PLL}}$ of GFL-VSC experiences a significant change from point $P_1$ to $P_2$, a rather limited change can be observed in $\delta_{\text{s}}$. Hence, the traditional dual-blind-based scheme that predicts the increase of $\delta_{\text{S}}$ based on the leftward movement of the impedance trajectory might not be effective for GFL-VSC systems.

*2) Dynamics of φ:* In this scenario, the dynamics of the angle $\varphi$, which is determined by the active/reactive power control, are further considered. Assuming that $Q_{\text{ref}} = kP_{\text{ref}}$ ($k$ is an arbitrary variable), and $k$ can be expressed as a tangent function of a certain angle $α_0$, i.e. $k=\tan α_0$. Based on (10) and (11), $I_{\text{dref}}$ and $I_{\text{qref}}$ can be further calculated as:

$$\begin{cases} I_{\text{dref}} = \dfrac{v_{\text{PCCd}} - kv_{\text{PCCq}}}{\sqrt{v_{\text{PCCd}}^2 + v_{\text{PCCq}}^2}} P_{\text{ref}} \\ I_{\text{qref}} = \dfrac{v_{\text{PCCq}} + kv_{\text{PCCd}}}{\sqrt{v_{\text{PCCd}}^2 + v_{\text{PCCq}}^2}} P_{\text{ref}} \end{cases} \quad (15)$$

Substituting (15) into (12), $\varphi$ can be simplified as:

$$\varphi = \arctan\left(\frac{I_{\text{qref}}}{I_{\text{dref}}}\right) = \arctan\left(\frac{v_{\text{PCCq}} + kv_{\text{PCCd}}}{v_{\text{PCCd}} - kv_{\text{PCCq}}}\right) \\ = \arctan\left(\frac{\tan(\delta_{\text{PCC}} - \delta_{\text{PLL}}) + \tan α_0}{1 - \tan α_0 \tan(\delta_{\text{PCC}} - \delta_{\text{PLL}})}\right) \quad (16) \\ = \delta_{\text{PCC}} - \delta_{\text{PLL}} + α_0$$

Substituting (16) into (14), $Z_{\text{PCC}}$ can be simplified and calculated as follows:

$$Z_{\text{PCC}}\angle\theta_{\text{ZP}} = \frac{E_{\text{R}}}{I_{\text{g}}} \angle -(\varphi + \delta_{\text{PLL}}) + Z_{\text{g}}\angle\theta_{\text{g}} \\ = \frac{E_{\text{R}}}{I_{\text{g}}} \angle -(\delta_{\text{PCC}} + \alpha_0) + Z_{\text{g}}\angle\theta_{\text{g}} \quad (17)$$

Comparing (14) and (17), one can observe that with the consideration of the open-loop active and reactive power control, $\varphi+\delta_{\text{PLL}}$ is equal to $\delta_{\text{PCC}}+\alpha_0$ as analyzed. Thus, the angle $\varphi$ could offset the influence of $\delta_{\text{PLL}}$ on $Z_{\text{PCC}}$ as shown in (17). With this influence, the trajectory of $Z_{\text{PCC}}$ is dependent on $\delta_{\text{PCC}}$ and $\alpha_0$. Based on (10) and (16), one can also observe that with further consideration of the outer power control loop, the angle difference $\delta_{\text{PCC}}-\delta_{\text{PLL}}$ could be involved from the dq-components of the PCC voltage ($v_{\text{PCCdq}}$) to the dq-components of the current references ($I_{\text{dqref}}$), and further reduce the dynamic influence of $\delta_{\text{PLL}}$ on the impedance trajectory owing to the opposite dynamic of $\delta_{\text{PLL}}$ based on (17). Consequently, when considering the active/reactive power control, the impedance trajectory is characterized by $\delta_{\text{PCC}}$ and $\alpha_0$, i.e., the arc-tangent value of $P_{\text{ref}}$ and $Q_{\text{ref}}$. In this situation, it can be even more misleading to predict the power swing based on $Z_{\text{PCC}}$, as seen from (17), $Z_{\text{PCC}}$ does not reflect the dynamics of $\delta_{\text{PLL}}$ at all.

In general, the impact of the GFL-VSC-control dynamic on the efficacy of PSB and OST can be summarized as follows:

- In GFL-VSC systems, it is the **control-dependent** angle $\varphi+\delta_{\text{PLL}}$ that characterizes the impedance trajectory and indicates the risk of LOS, referring to (14), which is fundamentally different from the power swing dynamics of SG that is characterized by the **physical** power angle difference $\delta_{\text{S}}$ between two power sources.
- With further consideration of the outer power control, the angle $\varphi$ can offset the influence of $\delta_{\text{PLL}}$ on $Z_{\text{PCC}}$, referring to (17). As a result, the power swing detection based on $Z_{\text{PCC}}$ is misleading, as $Z_{\text{PCC}}$ is independent of $\delta_{\text{PLL}}$. This influence makes it difficult to detect stable and unstable power swings by merely using the $Z_{\text{PCC}}$ trajectory, and may even lead to the maloperation of the conventional two-blinder scheme-based PSB and OST functions.

These influences on the efficacy of PSB and OST will be validated by simulation case studies in Section IV.

## IV. CASE STUDIES

### A. System Descriptions

To analyze the efficacy of PSB and OST functions in GFL-VSC systems, time-domain simulations are performed in the PSCAD/EMTDC platform, with the system depicted in Fig. 11. It is worth noting that the grid impedance is considered as a pure inductance. $L_{\text{S}}$ represents the equivalent inductor of the filter, while $L_1$ and $L_2$ represent the different line inductors, respectively. $RL_1$-$RL_4$ represent distance protection relays. The key parameters of the system are given in Table I.

To investigate the influence of the outer-loop controller and the angle $\varphi$, two different control modes are considered, which are illustrated in Fig. 12. Fig. 12 (a) shows the constant power factor (CPF) mode with no active and reactive power control loop, where the angle $\varphi$ remains constant during power swing situations. Fig. 12 (b) shows the open-loop power control (OPC) mode, where the open-loop active and reactive power control shown in (11) is used, and the dynamic of angle $\varphi$ influences the power swing. During normal operation, in both two control modes, the current reference selection is switched to 1. Owing to the fixed current references $I^*_{\text{dref}}$ and $I^*_{\text{qref}}$ in the CPF mode, the phase angle $\varphi$ remains constant. However, in the OPC mode, the current references are dictated by the active and reactive power controllers given in (15), and the phase angle $\varphi$ may vary during power swing situations. During the fault conditions, the current reference selection is changed to 2 [21], [22], and both CPF and OPC modes generate current references according to the grid codes depicted in Fig. 6.

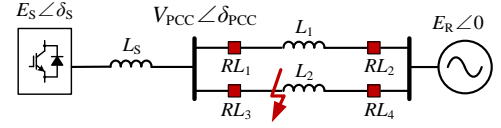

Fig. 11. Single-line diagram of the test GFL-VSC system.

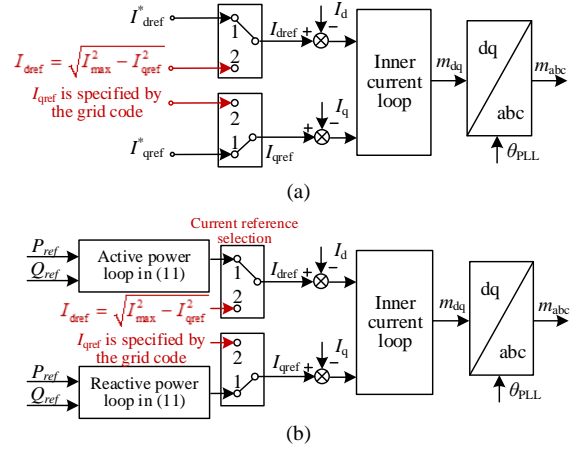

Fig. 12. Outer loop control diagram: (a) Constant power factor (CPF) mode, (b) Open-loop power control (OPC) mode.

### B. Maloperation of PSB and OST Functions in CPF Mode

Since the impedance trajectory of the GFL-VSC system is characterized by the angle $\delta_{\text{PLL}}$ and $\varphi$ according to (14), which may lead to the maloperation of PSB and OST functions with the two-blinder scheme. For simplicity, the CPF mode with $\varphi=0°$ is considered to see the maloperation of PSB and OST functions. To generate the power swing, a three-phase fault occurs at $t$=30s on the transmission line 2 near the relay $RL_3$ (see Fig. 11), and the fault is cleared after 0.3s by breaking the relays $RL_3$ and $RL_4$ in Fig. 11. The power swing is monitored on $RL_1$ in Fig. 11, and this case is studied with the CPF mode. The PSB and OST settings are referred to [2] and [4], and the phase angle of the Outer, Middle, and Inner blinders, i.e. $\delta_{\text{O}}$, $\delta_{\text{M}}$, and $\delta_{\text{I}}$ in Fig. 3 are set to 90°, 100°, and 120°, respectively. Thus, the resistances for the Outer, Middle, and Inner blinders (i.e., $R_{\text{O}}$, $R_{\text{M}}$, and $R_{\text{I}}$) can be calculated as 0.72Ω 0.60Ω, and 0.42 Ω, respectively. The time intervals for PSB and OST (i.e., $\Delta T_{\text{PSB}}$ and $\Delta T_{\text{OST}}$) are calculated as 1.5 and 2.5 cycles according to (4) and (5), respectively [2], [4], and the main parameters can be given in Table II.

The impedance trajectory with the dual-blinder scheme is depicted in Fig. 13, while the trigger signals of the PSB and OST functions are shown in Fig. 14. The phase angle curves of $\delta_{PLL}$ and $\delta_S$ are shown in Fig. 15 (a) and (b), respectively.

TABLE I
MAIN PARAMETERS OF THE GFL-VSC MODEL

| Symbol | Descriptions | Values (p.u.) |
|---|---|---|
| $E_R$ | The L-L AC voltage rating | 220 kV (1 p.u.) |
| $P_{base}$ | The power rating of the VSC | 3025MW (1 p.u.) |
| $Z_{base}$ | Rated impedance | 16 Ω (1 p.u.) |
| $f_g$ | Grid frequency | 50Hz (1 p.u.) |
| $f_{lim}$ | The frequency limitation of PLL | ±5 Hz |
| $K_{pP}$ | Proportional gain of the PLL | 0.57 p.u. |
| $K_{iP}$ | Integral gain of the PLL | 0.0616 p.u. |
| $k_{pi}$ | Proportional gain of the inner loop | 1.2479 p.u. |
| $k_{ii}$ | Internal gain of the inner loop | 435.78 p.u. |

TABLE II
LINE PARAMETERS AND SETTINGS OF PSB AND OST

| Symbols | Descriptions | Values |
|---|---|---|
| $L_s$ | The inductance of the filter | 10 mH (0.2 p.u.) |
| $L_1$ | The inductance of Line 1 | 35.5 mH (0.7 p.u.) |
| $L_2$ | The inductance of Line 2 | 2 mH (0.04 p.u.) |
| $I^*_{dref}$ | The d-axis current reference | 1 p.u. |
| $I^*_{qref}$ | The q-axis current reference | 0 p.u. |
| $R_O$ | Resistance of the Outer Blinder | 7.2 Ω (0.45 p.u.) |
| $R_M$ | Resistance of the Middle Blinder | 6.0 Ω (0.38 p.u.) |
| $R_I$ | Resistance of the Inner Blinder | 4.2 Ω (0.26 p.u.) |
| $\Delta T_{PSB}$ | The time interval for PSB | 1.5 cycles |
| $\Delta T_{OST}$ | The time interval for OST | 2.5 cycles |

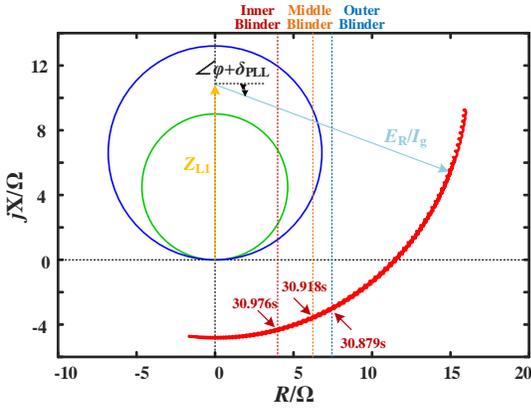

Fig. 13. Impedance trajectory with the dual-blinder scheme in GFL-VSC.

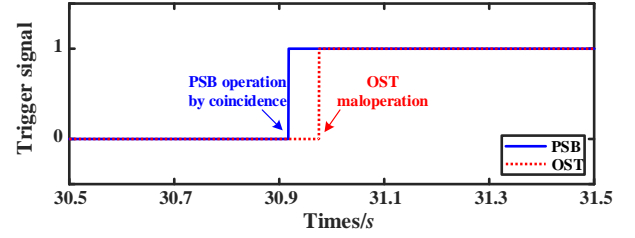

Fig. 14. Trigger signals of PSB and OST functions.

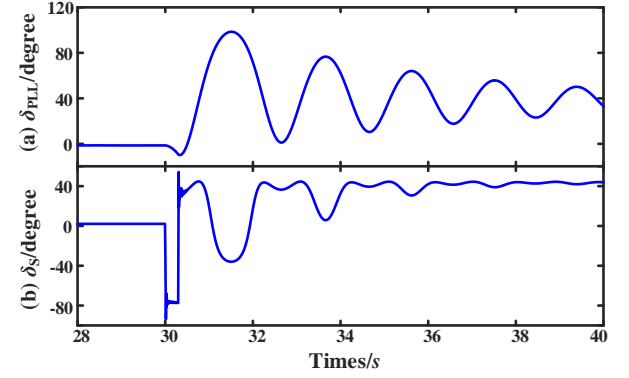

Fig. 15. Different angle curves: (a) $\delta_{PLL}$, (b) $\delta_S$.

It can be observed in Fig. 13 that the impedance trajectory crosses the Outer and Middle blinders with a time interval of $\Delta t_1$=30.918-30.879=0.039$s$, which is greater than the time interval set for the PSB function, i.e. $\Delta T_{PSB}$=1.5cycles=0.03$s$. It is worth noting that the PSB function is triggered at $t$=30.918$s$ purely by coincidence as shown in Fig. 14, as the impedance trajectory in the GFL-VSC system is characterized by $\delta_{PLL}$, and its movement can no longer accurately predict the changing of $\delta_S$.

Besides, the impedance trajectory crosses the Outer and Inner blinders with a time interval of $\Delta t_2$ = 30.976-30.879 = 0.097$s$, which is larger than $\Delta T_{OST}$=2.5 cycles=0.05$s$, leading to a maloperation of the OST function at $t$=30.976$s$ even with a stable power swing, as shown in Fig. 14.

The simulation results of different angle curves are shown in Fig. 15. It can be found that the maximum value of $\delta_{PLL}$ increases up to 95°, yet the angle $\delta_S$ does not increase this case, the power swing is characterized by the angle $\delta_{PLL}$, yet, $\delta_{PLL}$ does not necessarily correlate to the power angle $\delta_S$.

In general, the results indicate that the impedance trajectory in the GFL-VSC system is characterized by the sum of $\delta_{PLL}+\varphi$ ($\varphi$=0° in this scenario) based on (14), rather than the physical power angle $\delta_S$. According to Fig. 13, the impedance trajectory moves to the third quadrant when $\delta_{PLL}$ increases to 95°, yet no obvious increase can be observed on $\delta_S$ and its maximum value is around 50° during the power swing. The results also confirm that the power swing in such a system does not necessarily correlate to $\delta_S$, and thus maloperations could occur in PSB and OST functions that are based on the rate of change of the impedance, as shown in Figs. 13 and 14.

*C. Stable Power Swings*

To gain further insight into the power swing characteristics in the GFL-VSC systems, this case study investigates the influences of different control modes on power swing detection. A three-phase fault occurs at $t$=30s on transmission line 2 near $RL_3$ in Fig. 11, and the fault is cleared after 0.3s by breaking the $RL_3$ and $RL_4$ on transmission line 2. The power swing is monitored with $RL_1$ in Fig. 11. For comparison, four different sets of parameters are employed in this case, as shown in Case I with CPF mode and Case II with the OPC mode in Table III. The impedance trajectories under different control modes and parameters are shown in Fig. 16.

The different phase angle curves under different control modes are compared in Fig. 17. The PSB and OST triggering signals are shown in Fig. 18.

*1) CPF mode:* With the CPF mode, the angle $\varphi$ is maintained as a constant value in normal operations, i.e., $\varphi$=0° in Case I-A and $\varphi$=-11.2° in Case I-B. As the different value of $\varphi$, it can be observed from Fig. 16 that there are some offsets in different impedance trajectories when $I^*_{qref}$ changes from 0 p.u. (green line in Fig. 16) to -0.2 p.u. (blue line in

TABLE III
DIFFERENT PARAMETERS WITH STABLE AND UNSTABLE POWER SWING SITUATIONS

| Parameter | CPF mode (stable) | | OPC mode (stable) | | CPF mode (unstable) | OPC mode (unstable) |
|---|---|---|---|---|---|---|
| | Case I-A | Case I-B | Case II-A | Case II-B | Case III | Case IV |
| $L_s$ | 10 mH (0.2 p.u.) | 10 mH (0.2 p.u.) | 10 mH (0.2 p.u.) | 10 mH (0.2 p.u.) | 10 mH (0.2 p.u.) | 10 mH (0.2 p.u.) |
| $L_1$ | 15 mH (0.3 p.u.) | 15 mH (0.3 p.u.) | 15 mH (0.3 p.u.) | 15 mH (0.3 p.u.) | 35 mH (0.69 p.u.) | 28 mH (0.55 p.u.) |
| $L_2$ | 2 mH (0.04 p.u.) | 2 mH (0.04 p.u.) | 2 mH (0.04 p.u.) | 2 mH (0.04 p.u.) | 11 mH (0.22 p.u.) | 2 mH (0.04 p.u.) |
| $I^*_{dref}$ | 1 p.u. | 1 p.u. | / | / | 1 p.u. | / |
| $I^*_{qref}$ | 0 p.u. | -0.2 p.u. | / | / | 0 p.u. | / |
| $P_{ref}$ | / | / | 1 p.u. | 1 p.u. | / | 1 p.u. |
| $Q_{ref}$ | / | / | 0 p.u. | -0.2 p.u. | / | -0.2 p.u. |
| $R_O$ | 3.93 Ω (0.25 p.u.) | 3.93 Ω (0.25 p.u.) | 3.93 Ω (0.25 p.u.) | 3.93 Ω (0.25 p.u.) | 7.12 Ω (0.45 p.u.) | 6 Ω (0.38 p.u.) |
| $R_M$ | 3.35 Ω (0.21 p.u.) | 3.35 Ω (0.21 p.u.) | 3.35 Ω (0.21 p.u.) | 3.35 Ω (0.21 p.u.) | 5.97 Ω (0.37 p.u.) | 5.03 Ω (0.31 p.u.) |
| $R_I$ | 2.27 Ω (0.14 p.u.) | 2.27 Ω (0.14 p.u.) | 2.27 Ω (0.14 p.u.) | 2.27 Ω (0.14 p.u.) | 4.11 Ω (0.26 p.u.) | 3.46 Ω (0.22 p.u.) |
| $\Delta T_{PSB}$ | 1.5 cycles | 1.5 cycles | 1.5 cycles | 1.5 cycles | 1.5 cycles | 1.5 cycles |
| $\Delta T_{OST}$ | 2.5 cycles | 2.5 cycles | 2.5 cycles | 2.5 cycles | 2.5 cycles | 2.5 cycles |

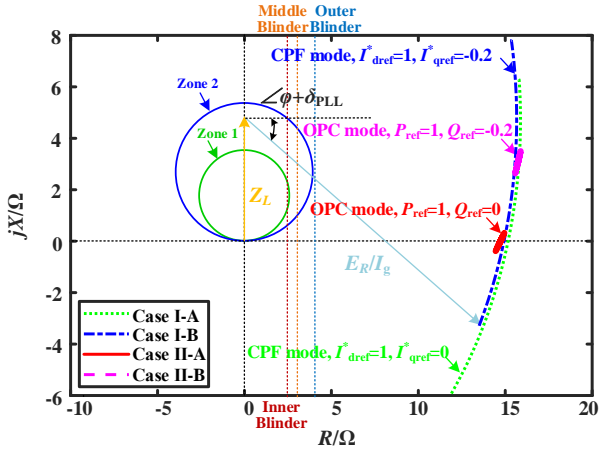

Fig. 16. Impedance trajectory with different modes and parameters.

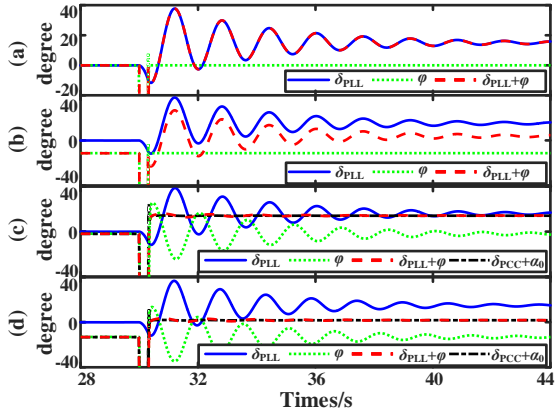

Fig. 17. The angle curves of $\delta_{PLL}$, $\varphi$, $\delta_{PLL}+\varphi$, and $\delta_{PCC}+\alpha_0$ in different situations: (a) Case I-A: $I^*_{dref}=1$, $I^*_{qref}=0$; (b) Case I-B: $I^*_{dref}=1$, $I^*_{qref}=-0.2$; (c) Case II-A: $P_{ref}=1$, $Q_{ref}=0$; (d) Case II-B: $P_{ref}=1$, $Q_{ref}=-0.2$.

Fig. 16) with the CPF mode. Moreover, as shown in Fig. 17 (a) and (b), the dynamic of the angle sum $\delta_{PLL}+\varphi$ is mainly relative to $\delta_{PLL}$ with CPF mode owing to the constant value of $\varphi$. Furthermore, one can observe that the impedance trajectories of Case I-A and Case I-B with the CPF mode do not cross any blinders in Fig. 16, which indicates that the PSB and OST functions are not triggered during these power swings, as shown in Fig. 18 (a) and (b). From these simulation results, it can be indicated that as the angle $\varphi$ remains constant, the dynamic of the impedance trajectory is only related to the dynamic of $\delta_{PLL}$ with CPF mode.

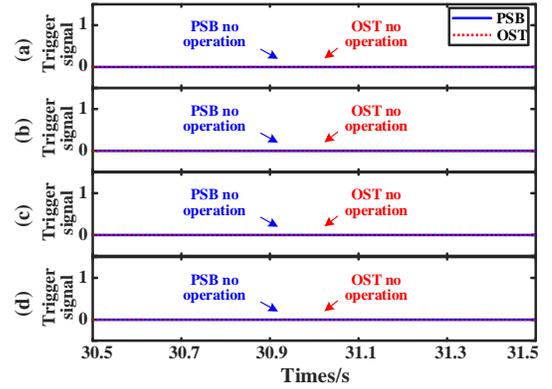

Fig. 18. The trigger signal of PSB and OST in different situations: (a) Case I-A: $I^*_{dref}=1$, $I^*_{qref}=0$; (b) Case I-B: $I^*_{dref}=1$, $I^*_{qref}=-0.2$; (c) Case II-A: $P_{ref}=1$, $Q_{ref}=0$; (d) Case II-B: $P_{ref}=1$, $Q_{ref}=-0.2$.

*2) OPC mode:* With the OPC mode, the angle $\varphi$ has an opposite direction and offsets the dynamic of $\delta_{PLL}$ based on (17), and the parameters are given as Case II-A and II-B in Table III. Similarly, when $Q_{ref}$ changes from 0 p.u. (red line in Fig. 16) to -0.2 p.u. (pink line in Fig. 16) with the OPC mode, the offset of the impedance trajectories in the impedance plane can be also observed. Compared with the impedance trajectories with the CPF mode, the impedance trajectories with the OPC mode have smaller moving ranges, this is consistent with the analysis in (17) that the dynamic influence of $\delta_{PLL}$ on the impedance trajectory can offset by $\varphi$, and the impedance trajectory is characterized by $\delta_{PCC}$ and $\alpha_0$. Furthermore, one can also observe from Fig. 17 (c) and (d) that with OPC mode, while there are larger but opposite dynamic dampings on both $\delta_{PLL}$ and $\varphi$, $\delta_{PLL}+\varphi$ is always equal to $\delta_{PCC}+\alpha_0$ with almost constant. The results from Fig. 17 (c) and (d) can also validate that as the opposite damping direction between $\varphi$ and $\delta_{PLL}$, the sum of $\delta_{PLL}+\varphi$ can be kept small, and the impedance trajectories with OPC mode in Fig. 16 can be kept in a very small range. Furthermore, one can observe that the impedance trajectories of Case II-A and II-B with the OPC mode do not cross any blinders in Fig. 16, which indicates that the PSB and OST functions are not triggered during these power swings, as shown in Fig. 18 (c) and (d). With further consideration of the outer power control in OPC mode, the angle $\varphi$ can offset the influence of $\delta_{PLL}$ on $Z_{PCC}$ referred to (17). As $Z_{PCC}$ is independent of $\delta_{PLL}$, it is more

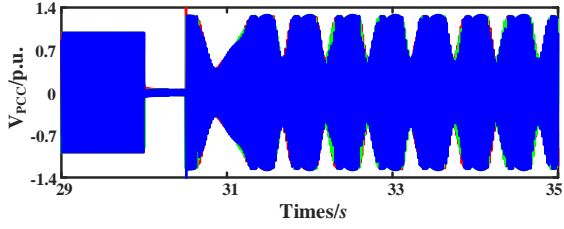

Fig. 19. Voltage curves of the PCC with Case III.

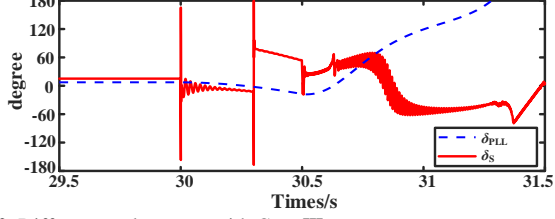

Fig. 20. Different angle curves with Case III.

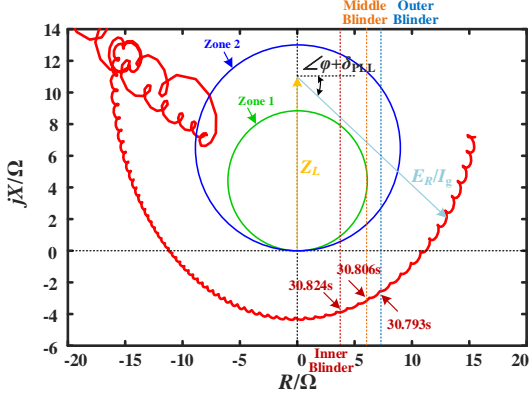

Fig. 21. Impedance trajectory with Case III.

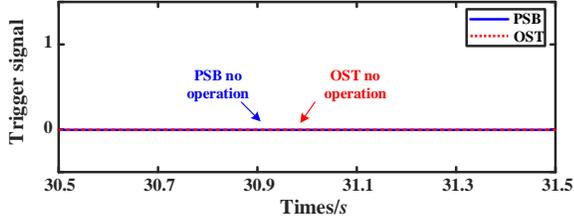

Fig. 22. Trigger signal of PSB and OST with Case III.

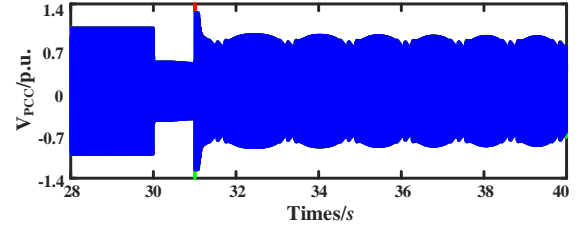

Fig. 23. Voltage curves of the PCC with Case IV.

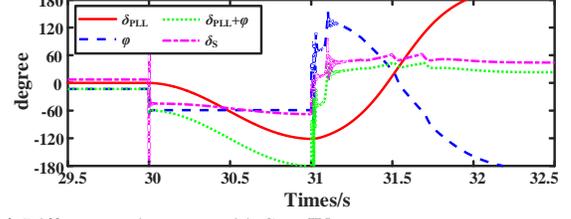

Fig. 24. Different angle curves with Case IV.

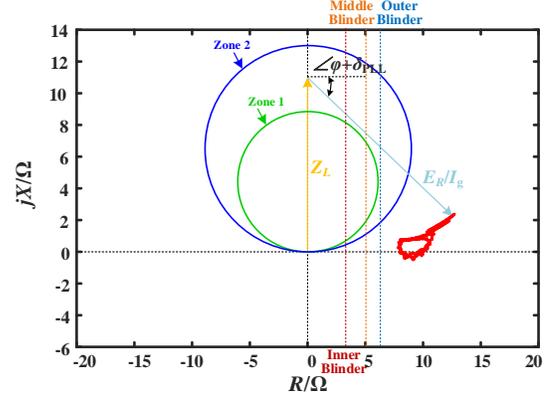

Fig. 25. Impedance trajectory with Case IV.

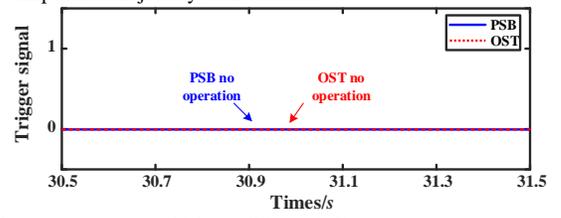

Fig. 26. Trigger signal of PSB and OST with Case IV.

difficult to detect power swing situations.

*1) CPF mode:* In this scenario, unstable power swing situations with the CPF mode are investigated. The key parameters of this scenario are shown in Cases III of Table III.

### D. Unstable Power Swings

Similarly, a three-phase fault occurs at $t$=30s on transmission line 2 near $RL_3$ in Fig. 11 with the CPF mode in this scenario, and the fault is cleared after 0.3s by breaking $RL_3$ and $RL_4$. The three-phase voltage curves are shown in Fig. 19, while different angle curves are shown in Fig. 20, The impedance trajectory with the unstable power swing is shown in Fig. 21, and the trigger signals of PSB and OST functions are shown in Fig. 22.

It can be observed in Figs. 19 and 20 that during the unstable power swing situation, as the angle $\varphi$ equals 0, the three-phase voltage curves finally lose synchronization. $\delta_{PLL}$ increases larger than 180 degrees and then diverges, and the divergence of $\delta_{PLL}$ occurs earlier than $\delta_S$. The simulation results indicate that the loss of synchronization in the GFL-VSC system is mainly due to the divergence of $\delta_{PLL}$, rather than the divergence of $\delta_S$. Furthermore, it can be also found from Fig. 21 that with the divergence of $\delta_{PLL}$, the impedance dramatically and its maximum value is around 50°. As $\varphi$=0 in trajectory moves to the left plane. The impedance trajectory crosses the Outer and Middle blinders with a time interval $\Delta t_1$=30.806-30.793=0.013$s$, and crosses the Outer and Inner blinders with a time interval $\Delta t_1$=30.824-30.793=0.031$s$, the maloperation of PSB and OST functions are triggered according to the settings of $\Delta T_{PSB}$ and $\Delta T_{OST}$ in Table II. The trigger signals of PSB and OST functions in Fig. 21 also indicated the maloperation of PSB and OST functions, and this unstable power swing can not be detected correctly in this situation.

*2) OPC mode:* Furthermore, in this scenario, unstable power swing situations with the OPC mode are investigated. With the main parameters in Case IV of Table III, a three-phase fault occurs at $t$=30s on the transmission line 2 close to the $RL_3$ with the OPC mode, and the fault is cleared after 1$s$ by breaking the relays $RL_3$ and $RL_4$ on the transmission line 2.

The three-phase voltage curves are shown in Fig. 23, while different angle curves are shown in Fig. 24. The impedance trajectory with the unstable power swing is depicted in Fig. 25, while the trigger signals of the PSB and OST functions are shown in Fig. 26.

It can be found from Fig. 23 that the flip frequency of this unstable power swing is about 1Hz. Besides, according to Fig. 24, $\delta_{PLL}$ increases larger than 180° and finally diverges after the fault is cleared, yet, $\varphi$ moves in the opposite direction with $\delta_{PLL}$, causing $\delta_{PLL}+\varphi$ in a small range (around 60°). It can be also observed from Fig. 24 that the angle difference between VSC and the equivalent source $E_R$ (i.e. $\delta_S$) could be limited to around 50°. With the sum of $\delta_{PLL}+\varphi$ limited to a small range, the impedance trajectory could also be limited, which can be depicted in Fig. 25. Moreover, because the impedance trajectory does not cross any blinders, the PSB and OST functions are not triggered during this unstable power swing in Fig. 26. which means that this unstable power swing can not be detected correctly.

In general, with further consideration of the outer power control in OPC mode, the phase angle $\varphi$ could offset the influence of $\delta_{PLL}$ during dynamic situations, and the impedance trajectory is characterized by $\delta_{PCC}+\alpha_0$. This influence limits the impedance trajectory to a small range, making it harder to detect unstable power swing situations.

## V. Conclusion

This paper presents a theoretical analysis of the efficacy of PSB and OST functions in GFL-VSC systems. The findings can be summarized as follows.

1) Theoretical analysis reveals that power swing dynamics in GFL-VSC systems are characterized by the control-dependent angles $\delta_{PLL}$ and $\varphi$. Consequently, the impedance trajectory forms a circular pattern in the complex plane. This is fundamentally distinct from the power swing dynamics in SG systems, where the physical power angle difference (i.e., $\delta_S$) between two power sources characterizes the power swing dynamics.

2) With further consideration of the outer power control, the phase angle $\varphi$ in GFL-VSC systems can offset the impact of $\delta_{PLL}$. The resulting impedance trajectory is characterized by $\delta_{PCC}+\alpha_0$, with $\alpha_0$ representing the arc-tangent value of $P_{ref}$ and $Q_{ref}$. Simulation results demonstrate that this offset effect makes it difficult to detect both stable and unstable power swings by measuring the impedance trajectory.

3) As the impedance trajectory is characterized by the control-dependent angle $\delta_{PLL}+\varphi$, the dual-blinder scheme, commonly used in SG-based systems to predict physical power angle dynamics using impedance trajectory, cannot be directly applied for power swing detection in GFL-VSC systems.

All of the analysis and funding have been elaborated theoretically and confirmed by time-domain simulations. It should be noted that the main contribution of this paper is to reveal the different power swing characteristics between GFL-VSC systems and the SG-based systems, and investigate the efficacy of the conventional dual-blinder-based PSB and OST functions. The effective power swing detection methods based on these analyses in GFL-VSC systems will be one of the future works in this study.